\documentstyle[pre,aps,epsfig,preprint]{revtex}
\begin{document}
\draft
\title{Vibrational and stochastic resonances in  two coupled \\ overdamped
anharmonic oscillators}
\author{V.M.~Gandhimathi$^a$, S.~Rajasekar$^{a}$, J.~Kurths$^b$}
\address{\vskip 5pt  $^{a}$ School of Physics,
          Bharathidasan University, \\
         Tiruchirapalli--620 024,
         Tamilnadu, India
\vskip 5 pt $^{b}$ Institut f\"{u}r Physik, Postdam Universit\"{a}t, \\
                   Am Neuen Palais 10, D-14469 Postdam, Germany}
\maketitle
 \vskip 375pt
 \hskip 50pt{ E-mail: S.~Rajasekar: rajasekar@physics.bdu.ac.in} \vskip 1 pt
\hskip 95pt {V.M.~Gandhimathi: vm$_{-}$gandhimathi@yahoo.co.in}
\vskip 1 pt
\hskip 95pt {J.~Kurths: jkurths@agnld.uni-potsdam.de}
\newpage
\vskip 10 pt
\begin{abstract}
We study the overdamped version of two coupled anharmonic oscillators under the influence of both
low- and high-frequency forces respectively and a Gaussian noise term added to one of the two state variables of the
system. The dynamics of the system is first studied in the presence of both
forces separately without noise. In the presence of only one of the forces, no resonance
behaviour is observed, however, hysteresis happens there. Then the influence
of the high-frequency force in the presence of a low-frequency, i.e. biharmonic forcing, is studied. Vibrational resonance is found to occur when the amplitude of the high-frequency force is varied. The resonance curve resembles a stochastic resonance-like curve. It is maximum at the value of $g$ at which the orbit lies in one well during one half of the drive cycle of the 
low-frequency force and in the other for the remaining half cycle. Vibrational resonance is characterized using the response amplitude and mean residence time. We show the occurrence of stochastic resonance behaviour in the overdamped system by replacing the high-frequency force by Gaussian noise. Similarities and differences between both types of resonance are presented.

\end{abstract}
 \vskip 10 pt
\pacs{{\emph{PACS}}: 02.50.-r; 05.40.-a; 05.45.-a  \\
\noindent{{\emph{Keywords}: Vibrational resonance; low-frequency force; high-frequency force;
 stochastic resonance; noise; mean residence time.
  }} }
 \newpage
\noindent {\bf{1. Introduction}}
 \vskip 10 pt
Stochastic resonance (SR) [1-4] occurs when a weak periodic signal in a
nonlinear system is amplified by a noise of appropriate strength. In bistable
systems, it has been shown that other types of external driving can mimic
the role of noise in the amplification of the signal. Such external drivings
include a chaotic signal [5] or a high-frequency periodic force [6]. In the
latter case, the system is driven by a biharmonic signal, consisting of a low-frequency and a 
high-frequency periodic force. Landa and McClintock studied the effect of high-frequency force. They found a resonance-like behaviour in the response of the
system at the frequency of the low-frequency periodic force, when the amplitude
of the high-frequency force passed through a critical value and is termed
as vibrational resonance (VR). After this seminal work VR
has been studied in a bistable oscillator [7], a Duffing oscillator [8], excitable
systems [9] and spatially extended systems [10]. Experimental evidences of VR in a vertical cavity surface emitting laser [11], simple electronic
circuit [9] and in an overdamped Duffing oscillator [12] have been reported.

\vskip 10 pt
The study of two-frequency signals is important in communication because usually
a low-frequency signal modulates a high-frequency carrier signal and is also
an object of interest in many other branches of physics and biology such as
lasers [13], acoustics [14], neuroscience [15] and ionosphere [16]. Practical
importance especially in medicine of a high-frequency force has been reported [17-20].
\vskip 10 pt

In this paper we study a multistable system, i.e. a more generalized one compared
to former investigations, under the influence of a biharmonic force and noise. This enables
to study VR as well as SR and compare their characteristic features in the system

\begin{mathletters}
\begin{eqnarray}
\dot {x} & = & a_1 x - b_1 x^3 + \gamma xy^2 + f \sin \omega t + g \sin \Omega t
               +  \eta(t), \\
\dot {y} & = & a_2 y - b_2 y^3 + \gamma x^2y.
\end{eqnarray}
\end{mathletters}
The terms $ f \sin \omega t $ and $ g \sin \Omega t $ are the low-frequency signal
of amplitude $f$ and high-frequency $(\Omega \gg \omega)$ signal of amplitude $g$.
System (1) is an overdamped version of two coupled anharmonic oscillators. Its
potential in the absence of damping and external drivings is given by

\begin{equation}
V(x,y)  =  -  \frac{a_1}{2}x^2 + \frac{b_1}{4}x^4 - \frac{a_2}{2}
    y^2 + \frac{b_2}{4}y^4 - \frac{\gamma}{2}x^2y^2.
\end{equation}
In a very recent work, Baxter and McKane [21] considered eqs (1) as
a model for competition between two species. We fix the parameters of the system
as $ a_1 = 1.0$, $a_2 = 1.1$, $b_1 = 1.0$, $b_2 = 1.0$, $\gamma=0.01$. For this
choice, the potential is a four-well potential. Figure (1) shows the 3-dimensional plot of the four-well potential. The potential wells are designated as $V_{++}$ for $x>0$, $y>0$; $V_{+-}$ for $x>0$, $y<0$; $V_{-+}$ for $x<0$, $y>0$;
$V_{--}$ for $x<0$, $y<0$. $\eta$ is additive white noise with $<\eta(t)  \eta(s)> = D^{2}  \delta(t-s)$ and $D$ is the noise intensity.

 \vskip 10 pt
The outline of the paper is as follows. In section 2 we discuss the response of
the system (1) in the presence of a low-frequency periodic force
only $(g = 0, D = 0)$ respectively a high-frequency periodic force only
$(f = 0, D = 0)$. In the presence of only one of
these forces, periodic orbits of the system (1) with a period equal to the single applied force
are found to exist. Chaos is not observed at all. We find that a 
cross-well orbit is realized above a critical amplitude of the
external driving. In section 3 we consider the noise free system
with both low- and high-frequency forces. We describe the
occurrence of VR and characterize it using the
response amplitude and mean residence time. In section 4 we show
the occurrence of SR in the presence of noise
and in the absence of high-frequency force. We discuss the
similarities and differences between VR and SR. Finally, section 5 contains our conclusions.
\vskip 10 pt
\noindent{\bf{2. Influence of only one periodic forcing}}
\vskip 10 pt
Before studying the system (1) in the presence of both, low- and 
high-frequency forces, we consider the noise free system with the low-frequency force alone $(g=0)$ respectively the high-frequency force alone
$(f=0)$. Though the potential has four wells, there is oscillating behaviour only in the wells $V_{++}$ and $V_{-+}$, while the other two wells $V_{+-}$ and $V_{--}$ are stationary, since the periodic
forces are added only to the $x$-component of the system (1). When the forces
and noise are included, a jumping of orbits between only the wells 
$V_{++}$ and $V_{-+}$ is observed for a range of control 
parameters considered in the present study. Therefore, we 
restrict ourselves to the motion confined to the wells $V_{++}$
and $V_{-+}$.
\vskip 10 pt 
We start with a low-frequency forcing. For a fixed $\omega$ and for 
small values of $f$, two periodic orbits one in the well $V_{++}$ 
and another one in the well $V_{-+}$ exist now. The size of the orbits 
is found to increase with increasing $f$. 
At a critical value $f_{\mathrm{c}}$ of $f$, after transient evolution the trajectory of the system resides in one of the
two potential wells, for example, say, $V_{++}$ for a certain interval of time and enters into the other well, say, $V_{-+}$ and stays there for sometime and the above process repeats. That is, the trajectory traverses both the potential wells. The $x$-component of the system changes from positive to negative and then to positive and so on whereas the sign of $y$ remains same. This 
type of jumping between two or more wells is termed as cross-well motion. For $\omega = 0.1$, we find $f_{\mathrm{c}} =
0.427$ as seen in the bifurcation diagram, Fig.2. In this and in the other bifurcation diagrams the ordinate represents the values of $x(t)$ collected at time $t$ equal to every integral multiples of $2 \pi / \omega$ (Poincar\'{e} points) after leaving sufficient transient evolution. Figure 2(a)
is obtained by varying the amplitude $f$ from a small value in the
forward direction. For the starting value of $f$, the initial
condition is chosen such that after some transient the orbit is
confined to the well $V_{++}$ only. For the other values of $f$,
the last state of the system corresponding to the previous value of $f$ is
chosen as the initial condition. In Fig.2(a) at $f = f_{\mathrm{c}}
= 0.427$, the $x$ value suddenly jumps to a negative value. Figure 2(b)
is obtained by varying $f$ in the reverse direction from the
value $2$. Different paths are followed in the Figs.2(a) and
2(b). That is, the system exhibits hysteresis when the control
parameter $f$ is varied smoothly from a small value to a large
one and then back to a small value. By numerical simulations we find
that $f_{\mathrm{c}}$ scales as $f_{\mathrm{c}}  = A \mathrm{e}^{0.75 \omega}$ + $B$  and is valid for $\omega$ in the range 0.01 to 1.0. The values of $A$ and $B$ are 0.754 and $-0.409$ respectively. That is, $f_{\mathrm{c}}$ increases with
increasing $\omega$.
\vskip 10 pt 
The bifurcation diagram and the Lyapunov exponents are calculated
for $f$ in the range [0,20]. Period doubling bifurcations
and chaotic dynamics are not observed here but there is only a stable periodic orbit of period $T
=2 \pi / \omega$.
\vskip 10 pt 
Next, we consider the system (1) in the presence of the
high-frequency force alone, that is $f = 0, D = 0$. The amplitude $g$
of the high-frequency force is varied in the forward as well as
reverse directions for $\Omega = 5$ and the bifurcation diagrams are plotted
(Fig.3) . Here again we find a cross-well
motion at a critical value $g_{\mathrm{c}} = 3.24$ of the force and a
hysteresis. Chaotic behaviour is again not observed. $g_{\mathrm{c}}$ is
numerically calculated for various values of $\Omega$ and is found
to increase with increasing $\Omega$. The calculated
$g_{\mathrm{c}}$ scales as $g_{\mathrm{c}} = C \Omega + D$ where
$C = 0.5$ and $D=0.298$, i.e., $g_{\mathrm{c}}$ varies
linearly with $\Omega$. $g_{\mathrm{c}}$  is valid for $\Omega$ in the range 0.01 to 10.

\vskip 10 pt
\noindent{\bf{3. Vibrational resonance}}
\vskip 10 pt
In the previous section, we studied the effect of the low-frequency force and
the high-frequency force separately in the system (1). Now, we consider the effect
of the high-frequency force on the response of the system in the presence of the
low-frequency force $( f \ne 0, g \ne 0)$ but in the noise-free case $(D=0)$ .
\vskip 10 pt
\noindent{\bf{A. Hysteresis}}
\vskip 10 pt
We fix $f = 0.2$, $ \omega = 0.1$ and $\Omega = 5$. For these values of the
parameters, in the absence of the high-frequency force $(g=0)$, we get $f_{\mathrm{c}} = 0.427$.
For $f=0.2$ in the absence of the high-frequency force there is no cross-well
motion, i.e. the low-frequency force alone is not sufficient to induce a cross-well motion.
We now study the response of the system by varying the amplitude $g$ of the high-frequency
force. For $g < g_{\mathrm{c}} = 2.67$ two periodic orbits with the same period $T =2 \pi /
\omega$ one in the well $V_{++}$ and another in the well $V_{-+}$ occur. For $g > g_{\mathrm{c}}$
the two periodic orbits merge and form a cross-well orbit. Figures 4(a) and
4(b) show the bifurcation diagrams obtained by varying $g$ in the forward and
reverse directions respectively. We can clearly notice a hysteresis. 
\vskip 10 pt
\noindent{\bf{B. Response amplitude $Q$ and mean residence time}}
\vskip10pt
In addition to the hysteresis, system (1) exhibits also the phenomenon of VR when $g$ is varied. To quantify the occurrence of VR, we use the response amplitude $Q$ of the system at the signal frequency $\omega$. It is defined as [6] $ Q = \sqrt{Q_{\mathrm{s}}^{2} + Q_{\mathrm{c}}^{2}}/f$ with
\begin{equation}
Q_{\mathrm{s}}   =  \frac {2} {nT} \int^{nT}_{0} x(t) \ \sin ( 2 \pi t / T) \ dt , \quad 
Q_{\mathrm{c}}   =  \frac {2} {nT} \int^{nT}_{0} x(t) \ \cos ( 2 \pi t / T) \ dt ,
\end{equation}
where $T$ is the period of the response and $n$ is an integer. We
numerically calculate $Q$ with a low-frequency force only, a 
high-frequency force only and with both forces. In the case of $f=0$, the response amplitude is measured as $ Q = \sqrt{Q_{\mathrm{s}}^{2} + Q_{\mathrm{c}}^{2}}$. $Q_{s}$ and $Q_{c}$ measure the coefficients of the Fourier sine and cosine components respectively of the output signal at the frequency $2 \pi/T$. $Q$ measures the amplitude of the response at the frequency $2 \pi/T$. When
the system is driven by only one force, the response $Q$
monotonically increases with the amplitude $f$ respectively $g$ of the corresponding driving force. This is shown in Fig.5(a) for $g=0, \omega =0.1$ and in Fig.5(b) for $f=0, \Omega =5.0$.
 Figure (6) shows a completely different result when both forces are
switched on. As $g$ increases, the resultant amplitude $Q$
increases and reaches a maximum value at $g = g_{\mathrm{max}} =
2.98$ but then decreases with further increase in $g$. This phenomenon is the VR, since the
occurrence is due to high-frequency force. We first describe both
qualitatively and quantitatively the VR and
then compare it with the well-known phenomenon of SR. \vskip 10 pt The
resonance curve in Fig.6 essentially consists of three regions: \vskip 1 pt
\hskip 75 pt Region-I   : $ 0 < g < g_{\mathrm{c}} (=2.67)$. \vskip 1 pt
\hskip 75 pt Region-II  : $ g_{\mathrm{c}}  \le g \le g_{\mathrm{max}} (=2.98)$. \vskip 1 pt
\hskip 75 pt Region-III : $ g > g_{\mathrm{max}}$. \vskip 1 pt

Figures (7) and (8) show the phase portraits and trajectory plots
for a few values of $g$ in the interval [0,6]. In all the three
regions the motion is periodic with the period $T =2 \pi /
\omega$. For  $ 0 < g < g_{\mathrm{c}} = 2.67$, that is in
region-I, each periodic orbit is confined only to one well. There
is no cross-well motion (cf. Figs.7(a-b) for $g=1.0$
and $2.5$). In region-I, the amplitude $Q$ increases smoothly and
slowly with $g$. At $g_{\mathrm{c}} = 2.67$, a cross-well motion is
initiated. Figures 7(c) and 7(d) show the coexistence of two such
orbits for $g=2.67$. In Fig.7(c) the period-$T$ orbit
spends relatively a very small amount of time in the region $x>0$,
that is in the well $V_{++}$. In contrast to this, the orbit in
Fig.7(d) traverses only a small region and relatively a very small
time in the well ${V_{-+}}$. In all subplots of Fig.7, the state
variable $y$ is always positive.
\vskip 10 pt
 We also calculate the mean residence time $\tau_{\mathrm{MR}}$ spend in the wells $V_{++}$ and
$V_{-+}$ by the orbit for $g$ in the interval $[g_{\mathrm{c}},6]$. In our numerical calculation of $\tau_{\mathrm{MR}}$ for $ g = g_{\mathrm{c}}$ we have chosen initial conditions such that the
orbit spending most of the time in the well $V_{++}$. Figure 9(a)
shows the plot of ln$(\tau_{\mathrm{MR}})$ versus $1/(g-g_{\mathrm{c}})$
in the well $V_{++}$ and Fig.9(b) in the well $V_{-+}$. The barrier heights for the path $V_{++} \rightarrow V_{-+}$  and $V_{-+} \rightarrow V_{++}$ are the same. At $g = g_{\mathrm{c}}$, $\tau_{\mathrm{MR}}$ of the orbit in the well $V_{-+}$ is
very small while for $V_{++}$ it is close to $T=2 \pi / \omega$.
As $g$ increases, $\tau_{\mathrm{MR}}$ in $V_{++}$ decreases rapidly, while
$\tau_{\mathrm{MR}}$ in $V_{-+}$ increases rapidly. As $ g \rightarrow g_{\mathrm{max}}
$ (which is denoted by $X$ in Fig.9), $ \tau_{\mathrm{MR}}$ in the well
$V_{++}$ decreases to $T/2$, while $\tau_{\mathrm{MR}}$ in $V_{-+}$
increases to $T/2$. At $g = g_{\mathrm{max}}$ the orbit lies in the well
$V_{++}$ during one half of the drive cycle of the lower frequency 
and in the well $V_{-+}$ in the other half of the cycle as shown 
in Figs.7(e) and 8(d). The above dynamics is in region-II. In this 
region, there is a rapid increase in the amplitude $Q$ and region-II is very narrow.
\vskip 10 pt
Though the period of the orbit in the region-III is still $T$, 
the form of the orbit is different from those in the regions I and II. 
Figure 7(f) depicts the phase portrait of the period-$T$ orbit for
$g = 6.0$. The corresponding trajectory plot is shown in Fig.8(f).
These two figures can be compared with the phase portrait and 
trajectory plot of orbits in the regions I and II. The orbit 
appears to move on a `$\infty$' shaped surface. When $g$ is increased 
above $g_{\mathrm{max}}$, the mean residence time of the orbit 
in the potential well $V_{++}$ and in the well $V_{-+}$ decreases.
This is evident from the trajectory plots, Fig.8(e) and 8(f).
In the region-III $Q$ decreases much slower than it increases
in region-II. From the above, we find that as $g$ is increased from
$g_{\mathrm{c}}$, $ \tau_{\mathrm{MR}}$ of the orbit in the well $V_{++}$ decreases from $T$, becomes $T/2$ at $g=g_{\mathrm{max}}$ and then decreases with further increase in $g$. For the well $V_{-+}$, when $g$ is increased from $g_{\mathrm{c}}$, the value of $ \tau_{\mathrm{MR}}$ increases from zero, reaches the maximal value $T/2$ at $g=g_{\mathrm{max}}$ and then decreases for $g>g_{\mathrm{max}}$.
\vskip 10 pt
We study the dependence of $Q$ on the frequencies $\omega$ and $\Omega$ of the driving forces. In Fig.10(a), $Q(g)$ is plotted for different values of $\omega$, namely $\omega = 0.1$, $0.25$, 
$0.5$ and with $\Omega =5$, $f=0.2$. With increasing $\omega$, $Q_{\mathrm{max}}$ decreases. The value of $g_{\mathrm{max}}$ (at which $Q$ is maximum) is also found to depend on $\omega$. The resonance peak becomes more and more sharp with decreasing $\omega$. In Fig.10(b) the resonance curve is plotted for different values of $\Omega$ for $\omega = 0.1$. $Q_{\mathrm{max}}$ is almost the same in all the cases. But $g_{\mathrm{max}}$ and the width of the resonance curve increases with $\Omega$.
\vskip 10 pt
We have studied the effect of high-frequency force with and without hysteresis phenomenon. For example, for $a_{1}=-1.0$, $a_{2} =-1.1$, $b_{1} =1.0$, $b_{2}=1.0$, $\gamma=0.01$, the potential of the system is a single-well and the bifurcation diagram for $f \in [0,50]$ show the absence of hysteresis behaviour. The parameter $g$ is varied for several fixed values of $f$ in the above specified interval for $\Omega=5.0$. In all the cases vibrational resonance is not observed. Recently, Ullner et al [9] analysed the effect of high-frequency forcing in excitable systems which have only one stable fixed point but perturbations above a certain threshold induce large excursions of the form of spikes in phase space. The duration of those excursions introduced an intrinsic time scale and the excitable systems exhibited vibrational resonance to the high-frequency harmonic driving.
\vskip 10 pt
The VR curve closely resembles the SR curve. In VR the role of noise is played in some sense by the high-frequency force. There are some characteristic 
similarities and differences between these two phenomena which will be discussed in the next section.
\vskip 10 pt
\noindent{\bf{4. Stochastic resonance}} \vskip 10 pt In this
section, we replace the high-frequency force in eq (1) by the noise term $
\eta (t)$ and show the occurrence of SR. Then we
make a brief comparison of VR with SR. Noise is added to the state variable $x$ as $x_{i+1}
\rightarrow x_{i+1} +  \zeta(t) $ with $ \zeta(t) = \sqrt {\Delta t} D \eta(t)$ after each
integration step $ \Delta t$. $\eta(t)$ represents Gaussian random
numbers with zero mean, unit variance and $D$ is the
intensity of the noise. In our numerical simulation we 
integrate the equation of motion with a time
step $ \Delta t = (2 \pi / \omega) / 2000$. The parameters in eqs (1) are
fixed as $ a_1 = 1.0$, $a_2 = 1.1$, $b_1 = 1.0$, $b_2 = 1.0$,
$\gamma = 0.01$, $f = 0.2$, $\omega = 0.1 $ and $g =0$.
 \vskip 10 pt
Figure 11(a) shows the numerically computed signal-to-noise ratio
$(SNR = 10 \log_{10}(S/N) \,\, \mathrm{dB})$ 
as a function of the noise intensity $D$.  Here $S$ and $N$ are the amplitudes of the signal peak and the noise background respectively. $S$ is read directly from the power spectrum
 at the frequency $\omega$ of the driving periodic force. To calculate
 the background of the power spectrum around $\omega$, we consider the power
 spectrum in the interval $\Omega  = \omega - \Delta \omega$ and
$\omega + \Delta \omega$ after subtracting the point namely
$\Omega = \omega$ representing the SR spike. The
average value of the power spectrum in the above interval is taken
as background noise level at $\omega$. In addition to the $SNR$ plot, the response amplitude $Q$ is measured by varying the noise intensity $D$. Figure 11(b) shows the variance of $Q$ against noise intensity $D$. Figure (12)
shows $x$ vs $t$ for four values of noise intensity $D$. The
dependence of $\tau_{\mathrm{MR}}$ on $D$ is depicted in Fig.13. These
three figures can be compared with the Figs.(6), (8) and (9) for VR. As in the $Q$
vs $g$ response for VR, we find in Fig.11(a) and (b) again three regions, here in dependence on
the noise strength $D$. The value of $D$ at which the response is maximum is found to be the same in both $SNR$ as well as $Q$ plot. 
 \vskip 10 pt
However there are also strong differences between VR and SR; mainly
to mention the following ones: In the $D$
region-I, similar to the region-I in Fig.6,  orbits confined
to a single-well alone exist. At $D = D_{\mathrm{c}} = 0.022$, 
a noise-induced cross-well motion is
initiated. At $D_{\mathrm{c}}$, $\tau_{\mathrm{MR}}$ in the well $
V_{++}$ is estimated as $700$. $\tau_{\mathrm{MR}}$ in the well $ V_{-+}$
is also $700$. This value of $\tau_{\mathrm{MR}}$ at $D_{\mathrm{c}}$ is
much higher than the period $T = 2 \pi / \omega$ of the driving force. In contrast,
for the high-frequency induced oscillatory dynamics (see Fig.9)
 $\tau_{\mathrm{MR}}$ for $V_{++}$ at $g_{\mathrm{c}}$ is $
\approx T$, while it is very small for $V_{-+}$. 
\vskip 10 pt
In the case of noise-induced dynamics, $\tau_{\mathrm{MR}}$ decreases for both
wells $V_{++}$ and $V_{-+}$ and is identical in both. But in the case of the
high-frequency induced dynamics, $\tau_{\mathrm{MR}}$ of $V_{++}$ decreases
from $T$, whereas that of $V_{-+}$ increases from a small value
with the amplitude $g$. In Fig.12 the maximum of $SNR$ is
realized for $D = D_{\mathrm{max}} = 0.32$ at which $ \tau_{\mathrm{MR}} =
T/2 $. In Fig.12(c), an almost periodic switching between the
positive and negative values of $x$ is clearly seen at $D_{\mathrm{max}}$. In Fig.6 \  $Q$ becomes maximum at
$g = g_{\mathrm{max}} = 2.98$ at which $ \tau_{\mathrm{MR}} = T/2$. In
region-III for the wells $V_{++}$ and $V_{-+}$, $\tau_{\mathrm{MR}}$
decreases with increase in $D$. For large noise the motion is
dominated by the noise term and the trajectory jumps erratically
between both wells. In the case of the high-frequency induced dynamics
also in region-III, for large values of amplitude $g$, the motion
is dominated by the high-frequency force which is evident in Fig.8(f). 
\vskip 10pt
Though the potential $V(x,y)$, eq (2), built up a four-well potential with the fixed parameters, only two of them determine the dynamics and are sufficient for the vibrational and stochastic resonance effects. This is because both the forces are added to the $x$-component of the system only. That is, the applied harmonic forces make only the potential wells $V_{++}$ and $V_{-+}$ to oscillate while the other two wells $V_{--}$ and $V_{+-}$ remain stationary. Consequently, the VR and SR effects are confined to the two wells $V_{++}$ and $V_{-+}$ only. The addition of external periodic forcings to both state variables induces transitions between all the wells. We note that eq (1) decouples into two single oscillators  for the parametric choices $a_2=a_1$, $b_2=b_1$, $\gamma=-3b_1$ under the change of variables $x=u+v$ and $y=u-v$. Only for these specific parametric choices it is enough to consider a one-dimensional double-well potential system instead of the system (1).  The values of parameters considered in our present study are different from the above special choice. Both the one-dimensional double-well potential system and two-dimensional four-well potential system are capable of showing VR and SR phenomena. However, the critical values of the parameters at which they occur will be different in both systems due to the coupling term. We cannot infer about the dynamics of the two-dimensional four-well potential system from the study of a one-dimensional double-well potential system.
\vskip 10pt
 \noindent{\bf{5. Conclusion}}
\vskip 10 pt
In this paper we have discussed the phenomenon of vibrational resonance in a
multistable system under the action of a two frequency signal, where one
frequency is much larger than the other. Adding the high-frequency force to the
system, which is already driven externally by the low-frequency one, results
in the amplification of the low-frequency signal. The response amplitude curve resembles
the SR profile. Therefore, we have also studied stochastic resonance by replacing the
high-frequency term with a Gaussian noise term. By comparing both VR and SR, the similarities and differences between both are studied. The major results of our present work are the following:\vskip 2 pt
(i) Vibrational resonance is found to occur only in the presence of hysteresis for the given choice of parameters. \vskip 2 pt
(ii) In both VR and SR, the maximum response is obtained only when the mean residence time is half of the period of the driving force $ f \sin \omega t$. \vskip 2 pt
(iii) The variation of the mean residence time with the amplitude of the 
high-frequency force in VR is found to be different from that of it with the amplitude of noise in SR. \vskip 2 pt
(iv) In the high-frequency force induced VR, the mean residence time $\tau_{\mathrm{MR}}$ in the well $V_{++}$ decreases from $T$, while there is an increase in $\tau_{\mathrm{MR}}$ in the well $V_{-+}$ from a small value. In contrast, in SR $\tau_{\mathrm{MR}}$ in the wells $V_{++}$ and $V_{-+}$ are identical and decreases with the increase in the amplitude of the noise. \vskip 2 pt
(v) The $SNR$ and the response amplitude $Q$ both peaks at the same value of $D$ at which $\tau_{\mathrm{MR}} =\pi / \omega$. This suggests that in addition to $SNR$ one can also use the quantity $Q$ to characterize SR.  \vskip 5 pt Recently, we have found that for the external force in the form of the modulus of a sine wave and a rectified sine wave, the maximum $SNR$ is obtained only when the sum of the residence times of both wells is $T/2$ [22]. Therefore, it is interesting to study in future VR with different periodic forces especially with the above two forces and generalize the concept of mean residence time at the maximum output. We expect experimental verifications in lasers, circuits etc, but also applications in ecology and biology.

\vskip 10pt
\noindent{\bf{Acknowledgement}}
 \vskip 10pt
The work of SR forms part of a Department of Science and
Technology, Government of India research project (VMG and SR). JK acknowledges the support 
from his Humboldt-CSIR research award. The authors acknowledge helpful suggestions of  anonymous referees.
\newpage
\begin{figure}
\begin{center}
\epsfig{figure=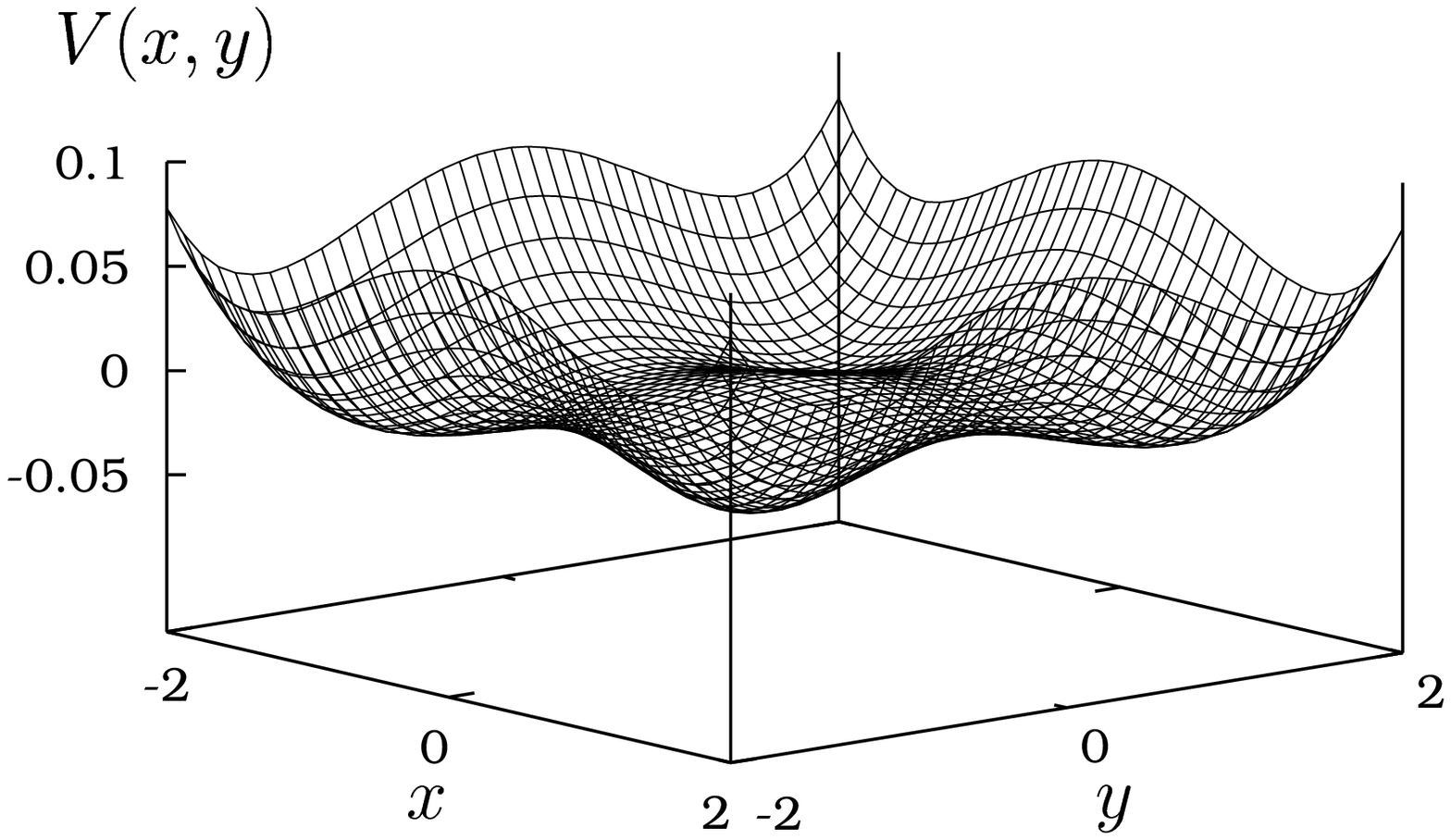, width=0.75\columnwidth}
\end{center}
\caption{3-dimensional plot of the potential $V(x,y)$ (eq.2) for the parameters $ a_1 = 1.0$, $a_2 = 1.1$, $b_1 = 1.0$, $b_2 = 1.0$, $\gamma=0.01$ } \label{Fig1}
\end{figure}

\begin{figure}
\begin{center}
\epsfig{figure=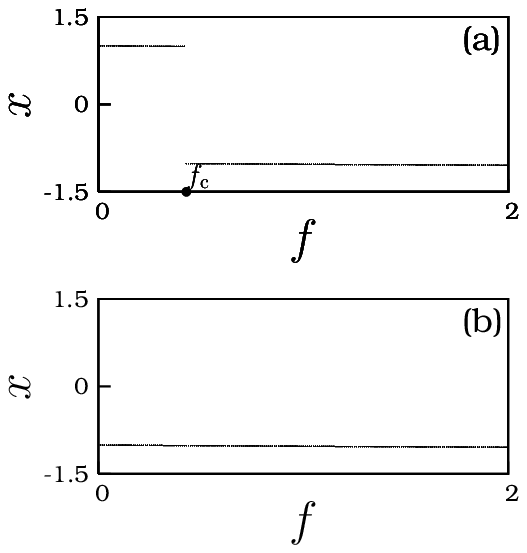, width=0.4\columnwidth}
\end{center}
\caption{Bifurcation diagrams in the presence of low-frequency force alone. $ \omega$ 
is set to 0.1. (a) $f$ is varied in the forward direction with the initial condition on the 
orbit lying in the well $V_{++}$ for the starting value of $f$. (b) $f$ is varied in the reverse direction from the value 2.} \label{Fig1}
\end{figure}
\newpage
\begin{figure}
\begin{center}
\epsfig{figure=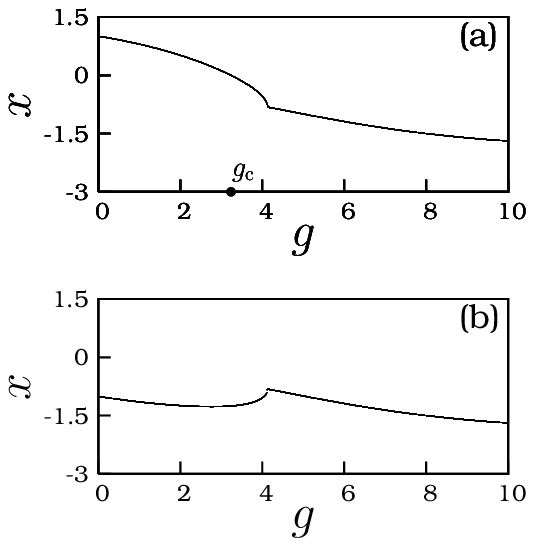, width=0.4\columnwidth}
\end{center}
\caption{Bifurcation diagrams in the presence of high-frequency force alone. $ \Omega$ 
is set to 5. (a) $g$ is varied in the forward direction with the initial condition on the 
orbit lying in the well $V_{++}$ for the starting value of $g$. (b) $g$ is varied in the reverse direction from the value 10. }
 \label{Fig2}
\end{figure}
\begin{figure}
\begin{center}
\epsfig{figure=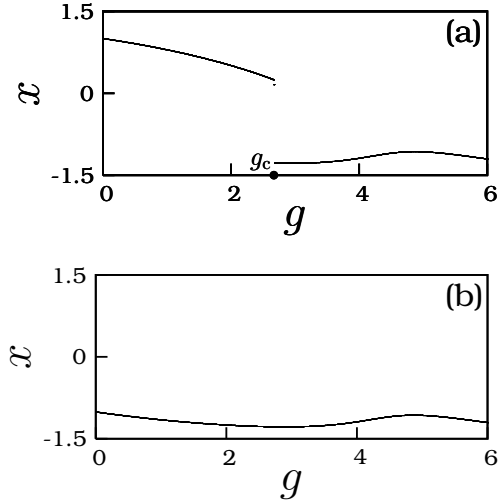, width=0.4\columnwidth}
\end{center}
\caption{Bifurcation diagrams in the presence of both low-frequency and 
high-frequency forces. Here $\omega = 0.1$, $\Omega = 5$ and $f=0.2$. $g$ is varied 
in the forward direction in (a) while it is decreased from a large
value to small value in (b).}
 \label{Fig3}
\end{figure}

\begin{figure}
\begin{center}
\epsfig{figure=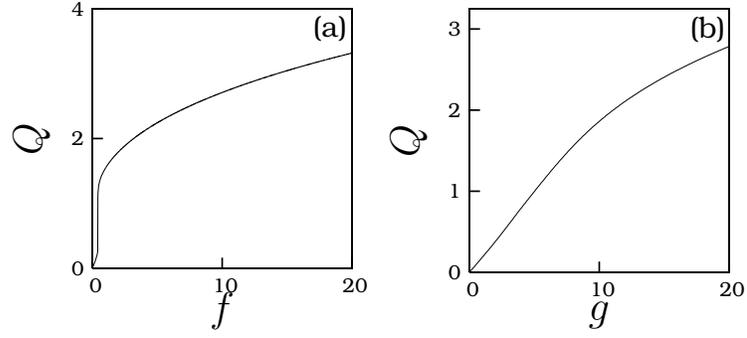, width=0.6\columnwidth}
\end{center}
\caption{(a) $Q$ versus $f$ where $g=0$ and $\omega=0.1$. 
(b) $Q$ versus $g$ where $f=0$ and $\Omega=5$.}
 \label{Fig4}
\end{figure}
\begin{figure}
\begin{center}
\epsfig{figure=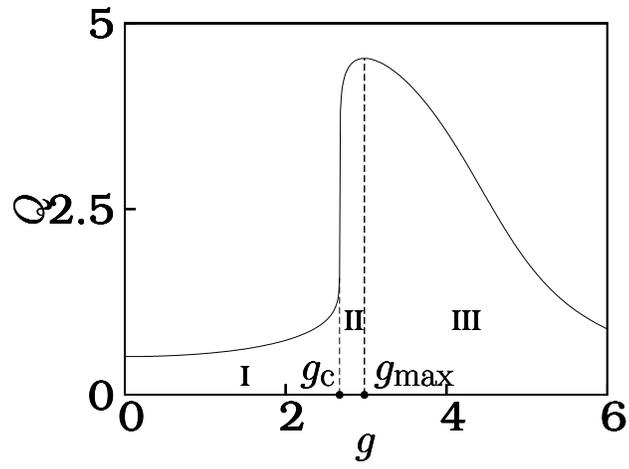, width=0.5\columnwidth}
\end{center}
\caption{The response amplitude $Q$ versus high-frequency amplitude for the 
parameters $f=0.2$, $\omega=0.1$ and $\Omega=5$.}
 \label{Fig5}
\end{figure}
\newpage
\begin{figure}
\begin{center}
\epsfig{figure=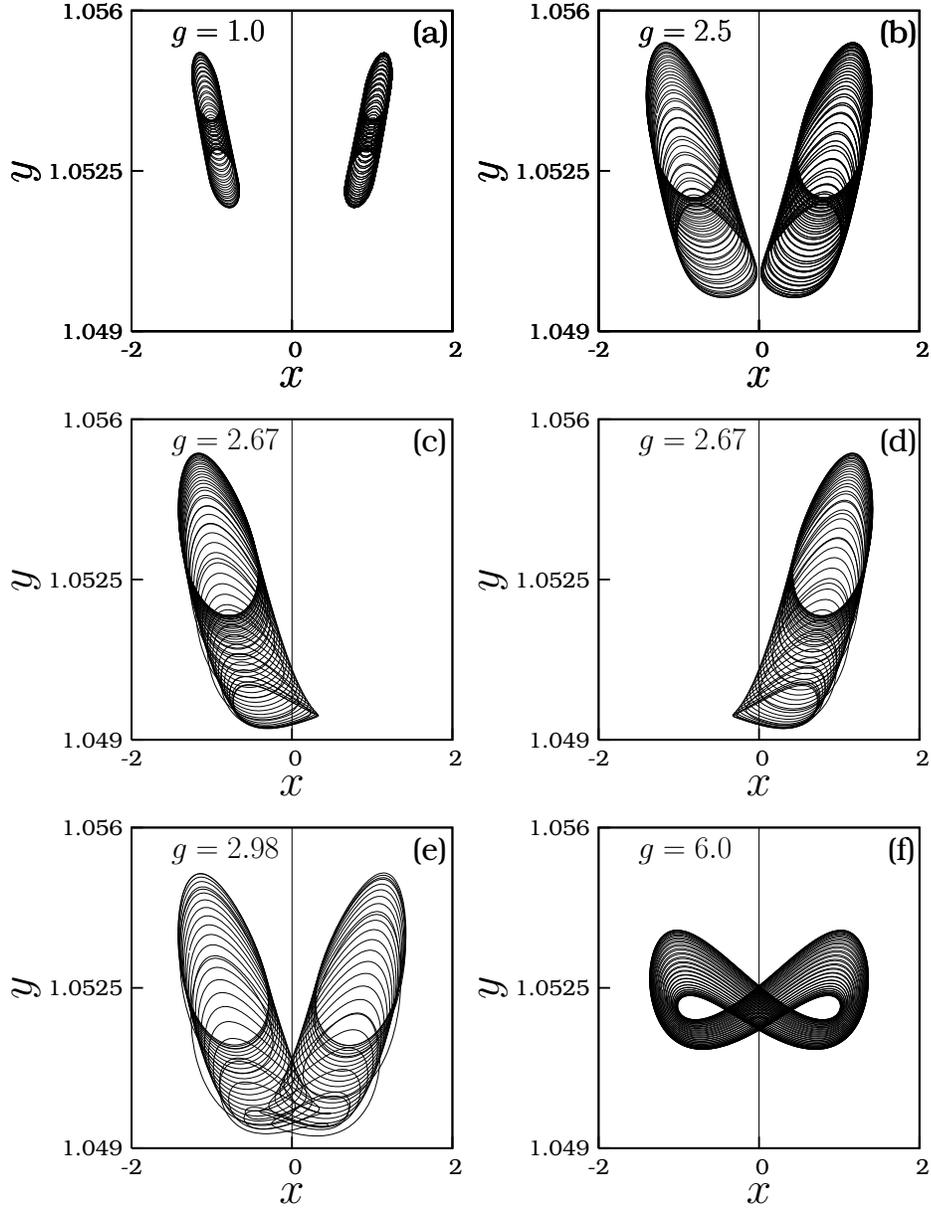, width=0.75\columnwidth}
\end{center}
\caption{Phase portrait for different values of high-frequency amplitude $g$.  
The other parameters are fixed at $f=0.2$, $\omega=0.1$ and $\Omega=5$.}
 \label{Fig6}
\end{figure}
\begin{figure}
\begin{center}
\epsfig{figure=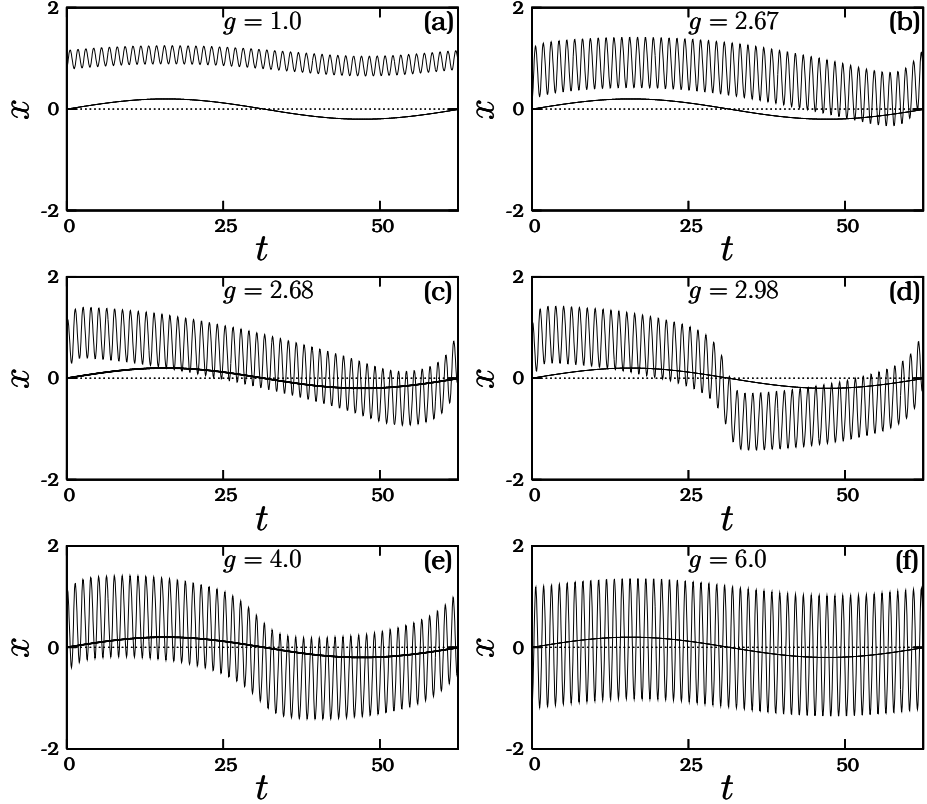, width=0.75\columnwidth}
\end{center}
\caption{Trajectory plots for few values of $g$. The parameters are $f=0.2$, $\omega=0.1$ and $\Omega=5$. The low-frequency force is also plotted.}
 \label{Fig7}
\end{figure}
\begin{figure}
\begin{center}
\epsfig{figure=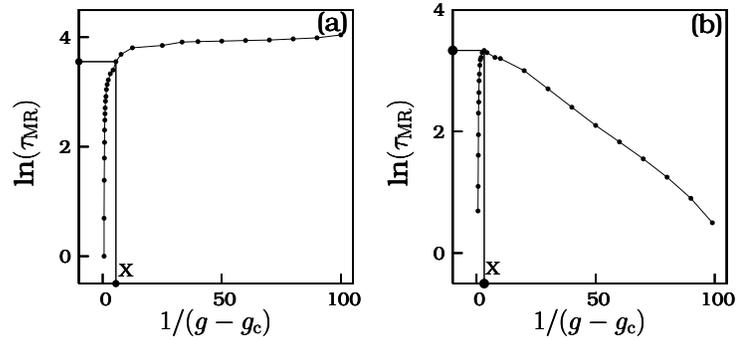, width=0.6\columnwidth}
\end{center}
\caption{Logarithmic plot of mean residence time of the orbit in the well $V_{++}$ (a)
and in the well $V_{-+}$ (b) versus $1/(g-g_{\mathrm{c}})$. }
 \label{Fig8}
\end{figure}
\begin{figure}
\begin{center}
\epsfig{figure=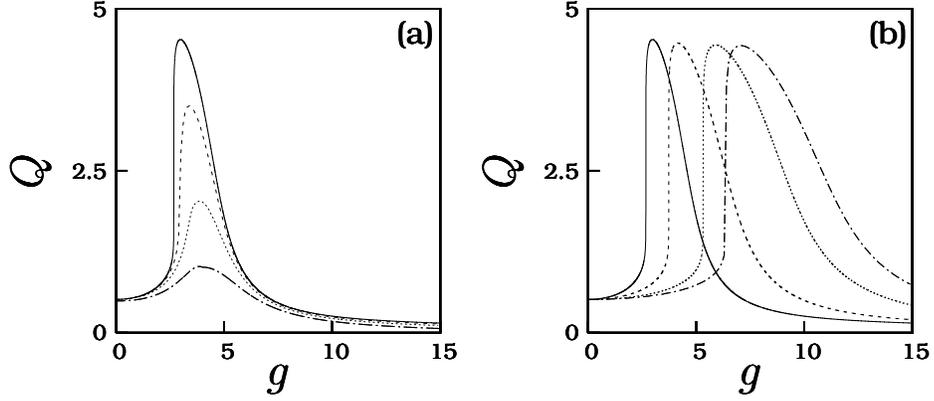, width=0.75\columnwidth}
\end{center}
\caption{ (a) Resonance curves for four fixed values of $\omega$. From top curve to
bottom curve the values of $\omega$ are $0.1$, $0.25$, $0.5$ and $1.0$ respectively. Here
$f=0.2$ and $\Omega=5$. (b) Resonance curves for four fixed values of $\Omega$ with $f=0.2$ and $\omega=0.1$. From left curve to right curve the values of $\Omega$ are $5$, $7$, $10$ and $12$ respectively.
}
 \label{Fig9}
\end{figure}
\begin{figure}
\begin{center}
\epsfig{figure=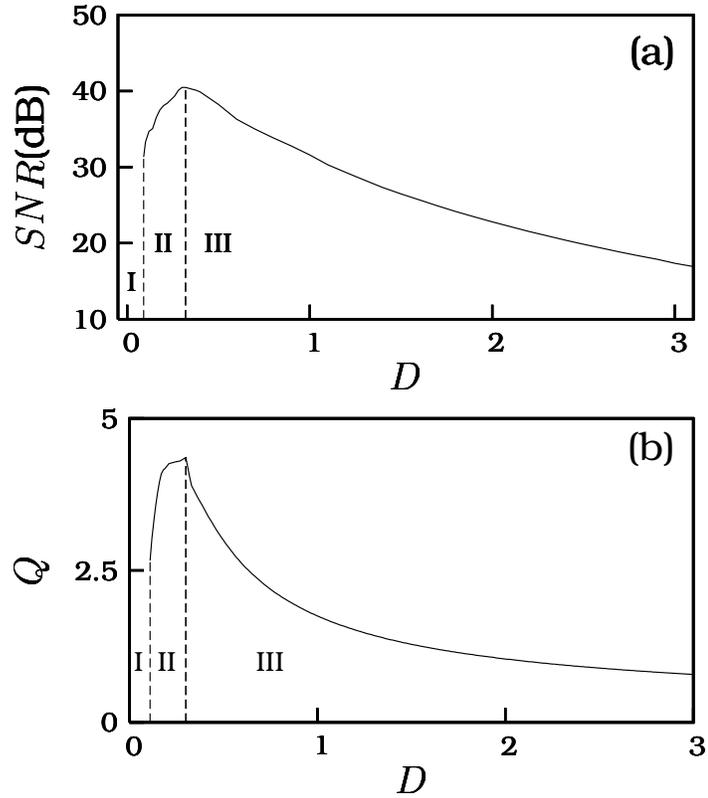, width=0.6\columnwidth}
\end{center}
\caption{(a) Signal-to-noise ratio plot and (b) Response amplitude plot for a range of noise intensity $D$ with $f=0.2$ and $\omega=0.1$.}
 \label{Fig10}
\end{figure}
\begin{figure}
\begin{center}
\epsfig{figure=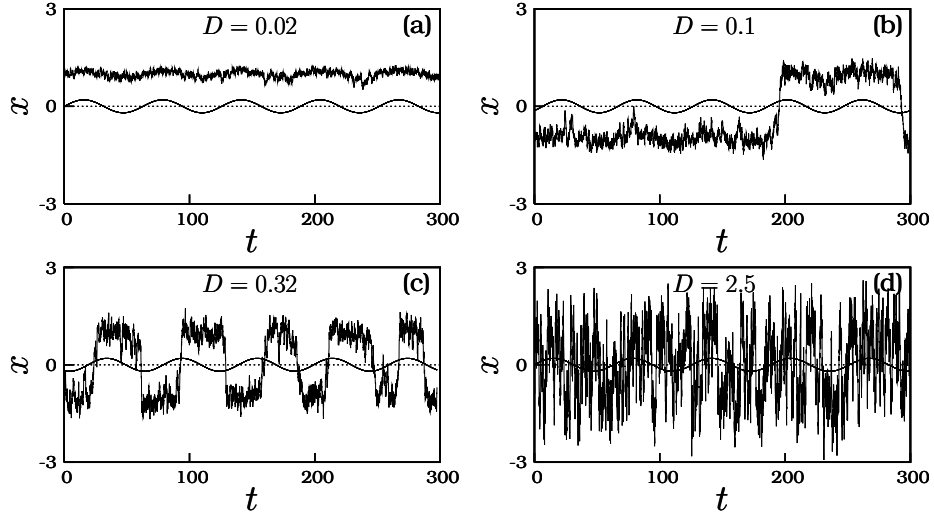, width=0.75\columnwidth}
\end{center}
\caption{Time series plot for few values of noise intensity $D$ with the system
parameters fixed at $f=0.2$, $\omega=0.1$.}
 \label{Fig11}
\end{figure}
\begin{figure}
\begin{center}
\epsfig{figure=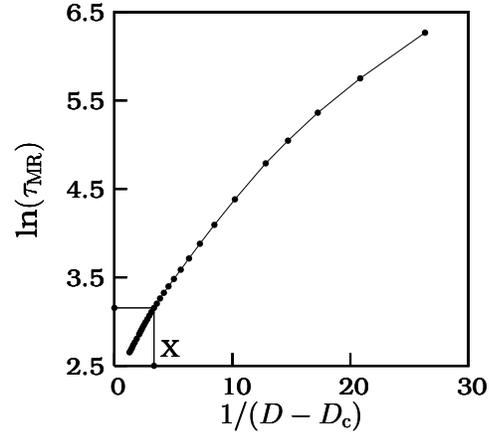, width=0.4\columnwidth}
\end{center}
\caption{ Logarithmic plot of mean residence time against the inverse of $(D-D_{\mathrm{c}})$. The system parameters are $f=0.2$ and $\omega=0.1$. $\tau_{MR}$ in $V_{++}$ and $V_{-+}$ are same.}
 \label{Fig12}
\end{figure}

\end{document}